# Comparative experimental and Density Functional Theory (*DFT*) study of the physical properties of $MgB_2$ and $AlB_2$

Devina Sharma[1,2], Jagdish Kumar[1,3], Arpita Vajpayee[1], Ranjan Kumar[2], P.K. Ahluwalia[1,3] and V.P.S. Awana[1,*]

[1]Quantum Phenomena and Applications Division, National Physical Laboratory, Dr. K.S. Krishnan Marg, New Delhi-110012, India
[2]Department of Physics, Punjab University, Chandigrah 160014, India
[3] Department of Physics, Himachal Pradesh University, Summerhill, Shimla-171005

In present study, we report an inter-comparison of various physical and electronic properties of $MgB_2$ and $AlB_2$. In particular the results of phase formation, resistivity $\rho(T)$, thermoelectric power $S(T)$, magnetization $M(T)$, heat capacity ($C_P$) and electronic band structure are reported. The original stretched hexagonal lattice with $a$ = 3.083 Å, and $c$ = 3.524 Å of $MgB_2$ shrinks in *c*-direction for $AlB_2$ with $a$ = 3.006 Å, and $c$ = 3.254 Å. The resistivity $\rho(T)$, thermoelectric power $S(T)$ and magnetization $M(T)$ measurements exhibited superconductivity at 39 K for $MgB_2$. Superconductivity is not observed for $AlB_2$. Interestingly, the sign of $S(T)$ is +ve for $MgB_2$ the same is −ve for $AlB_2$. This is consistent our band structure plots. We fitted the experimental specific heat of $MgB_2$ to Debye Einstein model and estimated the value of Debye temperature ($\Theta_D$) and Sommerfeld constant ($\gamma$) for electronic specific heat. Further, from $\gamma$ the electronic density of states (*DOS*) at Fermi level $N(E_F)$ is calculated. From the ratio of experimental $N(E_F)$ and the one being calculated from *DFT*, we obtained value of $\lambda$ to be 1.84, thus placing $MgB_2$ in the strong coupling *BCS* category. The electronic specific heat of $MgB_2$ is also fitted below $T_c$ using $\alpha$-model and found that it is a two gap superconductor. The calculated values of two gaps are in good agreement with earlier reports. Our results clearly demonstrate that the superconductivity of $MgB_2$ is due to very large phonon contribution from its stretched lattice. The same two effects are obviously missing in $AlB_2$ and hence it is not superconducting. *DFT* calculations demonstrated that for $MgB_2$ the majority of states come from σ and π 2p states of boron on the other hand σ band at Fermi level for $AlB_2$ is absent. This leads to a weak electron phonon coupling and also to hole deficiency as π bands are known to be of electron type and hence obviously the $AlB_2$ is not superconducting. The *DFT* calculations are consistent with the measured physical properties of the studied borides, i.e., $MgB_2$ and $AlB_2$

[*]Corresponding Author: Dr. V.P.S. Awana, Fax No. 0091-11-45609310: Phone no. 0091-11-45609357
e-mail-awana@mail.nplindia.ernet.in: www.freewebs.com/vpsawana/

# I. INTRODUCTION

Since the discovery of high temperature superconductivity (*HTSc*) in early 1987 [1], the fundamental physics discussion on electron – phonon interactions mediated so-called *BCS* type superconductors became only scant. However, the situation changed after the invention of superconductivity at 39 K in $MgB_2$ [2]. It was soon realized that unlike as for mysterious *HTSc* compounds, the mechanism of superconductivity in $MgB_2$ could yet be *BCS* type within strong coupling limits [3]. This was perceived from the fact that both Mg and B are very light elements and hence lattice contributions could be strong to the electron-phonon interactions, resulting in higher superconducting transition and still explainable within *BCS* limit [4]. Besides the light elements Mg and B in its formula, the lattice of $MgB_2$ is stretched in *c*-direction in comparison to other same structure borides viz., $TaB_2$, $AlB_2$, $ZrB_2$ or $MoB_2$ etc. [5-7]. Stretched lattice may result in instability and hence further more contribution to phonon interactions. It seems the basic stretched lattice structure of $MgB_2$ being constructed from relatively lighter elements Mg and B is responsible for strong electron phonon interactions. $MgB_2$ possess simple hexagonal $AlB_2$-type structure with space group *P6/mmm*. It contains the graphite-type boron layers, which are separated by hexagonal close-packed layers of magnesium. The magnesium atoms are located at the center of hexagons formed by boron. Also the distance between the boron planes is significantly large than in plane B-B length due to stretching of lattice in *c*-parameter (*c/a* ~ 1.14) in comparison to other borides viz., $AlB_2$ (*c/a* ~ 1.06) [8]. Its high superconducting transition temperature ($T_c$) comes from the exceptionally high vibrational energies in the graphite like boron planes and thus $MgB_2$ appears to follow conventional models of superconductivity [8-10].

As far as the electronic structure of $MgB_2$ is concerned the same has received much attention since invention of its superconductivity at a high temperature of 39 K [2]. There are many results that suggest holes in σ B-2*p* band via which electron-phonon coupling plays important roles in the superconductivity of $MgB_2$ [10-13].

Although the band structure calculations [10-13] and the strong Boron isotope effect [4], indicated $MgB_2$ to be a strong *BCS* type superconductor, the heat capacity reports are contradictory [14-18]. The electron phonon coupling constant being determined from $C_P(T)$ experiments varies from 0.7 (moderate coupling) [14] to 2.0 (strong coupling) [18] and intermediate as well [15-17]. This issue need to be probed.

In present study, we report an inter-comparison of various physical properties (transport, magnetic and thermal) and study the electronic properties of $MgB_2$ and $AlB_2$ using density functional theory. Further, we studied in detail the $C_P(T)$ of $MgB_2$ in high fields of up to 140kOe. Contrary to most earlier reports [14-17], but in agreement to one [18], our $Cp(T)$ results in combination with band structure calculations clearly show that $MgB_2$ is a strongly coupled *BCS* superconductor. For both the di-borides spin up and spin down electronic density of state (*DOS*) overlapped exactly indicating non-magnetic nature, as is observed experimentally. Further, absence of σ band at Fermi level for $AlB_2$ leads to a weak electron phonon coupling and hole deficiency due to dominating electron type π bands. This leads to the non superconducting behaviour of $AlB_2$.

## II. EXPERIMENTAL DETAILS

Polycrystalline samples of $MgB_2$ and $AlB_2$ are synthesized by solid-state route with ingredients of high purity Mg, B and Al. The nominal weighed samples are ground thoroughly, palletized, encapsulated in soft iron tube and inserted in separate programmable furnaces under flow of argon at one atmosphere pressure. The temperature of furnaces is programmed to reach $850^oC$ and $900^oC$ with heating rate of $10^oC$ per minute for $MgB_2$ and $AlB_2$ respectively. These temperatures are hold for two and half hours and subsequently cooled to room temperature under flow of argon atmosphere. The X-ray diffraction (*XRD*) pattern of the compound was recorded with a diffractometer using $CuK_\alpha$ radiation. Resistivity measurement $\rho(T)$, *DC* magnetic susceptibility in both zero-field-cooled (*zfc*) and field-cooled (*fc*) situations along with *AC* susceptibility are carried out on a physical property measurement system (*PPMS*) of Quantum Design (*QD*). The heat capacity $C_p(T)$ measurements under magnetic field of up to 14 Tesla are also done on *QD-PPMS*.

## III. CALCULATION DETAILS

### (a) Computational parameters

The lattice parameters were obtained by SIESTA (Spanish Initiative fo Electronic Simulation on Thousands of Atoms) code that implements density functional theory using localized atomic orbitals in terms of multiple zeta functions. We used double zeta polarised (DZP) basis set. The pseudopotentials were generated by atom program as included in SIESTA pacage using GGA by PBEsol exchange correlation functional. The transferability of pseudopotentials of Al, Mg and B were tested by reproducing corresponding bulk properties. The Brillouin zone integrations were performed using 15×15×15 special k-grid Monkhrost pack for lattice parameters and 20×20×20 for final properties. The band structure for these compounds were calculated by using full-potential linear augmented plane wave method (FP-LAPW) implemented in ELK code by taking calculated lattice parameters. The properties like electronic density of states (DOS) were calculated using both the codes and matches well with each other.

### (b) Methodology

The studied diborides consists of hexagonal planes of boron atoms separated by planes of Mg/Al atoms. The unit cell of their hexagonal structure has one Mg atom at (0, 0, 0) and two boron atoms at (1/3, 2/3, 1/2) and (2/3, 1/3, 1/2) respectively. For determination of lattice constant we fixed c/a at different values and then calculated total energy by varying a around our experimentally obtained values by keeping fractional coordinates of atoms fixed. Then from the plots of total energy versus '*a*' at different '*c/a*' we obtained the structural ground state with minimum energy. This gave us calculated lattice parameters *c/a* and *a*, that are in good agreement with our experimental measurements and other reports. By taking these calculated parameters we have we relaxed the atomic positions using conjugate gradient method to a maximum force tolerance of 0.1eV/Å. The relaxed atomic positions were in good agreement with experimental values. Then by taking these final values of lattice parameters and atomic positions we have calculated the other properties like DOS and band structure for both these compounds.

## III. RESULTS AND DISCUSSION

Room temperature X-ray diffraction (*XRD*) patterns for both $MgB_2$ and $AlB_2$ samples are shown in Fig. 1 along with their Reitveld refinement. Both samples crystallize in simple hexagonal $AlB_2$-type structure with space group *P*6/*mmm*. Though the structure (hexagonal) and space group *P*6/*mmm* remain the same for $AlB_2$, all the *XRD* peaks positions are shifted towards higher angle side for $AlB_2$ in comparison to peak positions of $MgB_2$, indicating a decrease in lattice parameters. The Rietveld refined lattice parameters are *a*=3.083Å, *c*=3.524Å for $MgB_2$ and *a*=3.006Å, *c*=3.254Å for $AlB_2$ sample. The *c/a* values are 1.14 and 1.06 respectively for $MgB_2$ and $AlB_2$. The structural information regarding $MgB_2$ and $AlB_2$ is in confirmation with the reported values [19-21]. The coordinate positions are Mg/Al at (0,0,0) and B at (1/3,2/3,1/2).

The resistivity $\rho(T)$ plots of polycrystalline $MgB_2$ and $AlB_2$ are shown in the Fig. 2. The critical temperature $T_c$ ($\rho \rightarrow 0$) and room temperature resistivity ($\rho^{300K}$) for $MgB_2$ are found to be ~ 38 K and 300 μΩ-cm respectively; while $AlB_2$ sample shows no superconducting transition. The experimental plots are fitted using power law, $\rho(T) = \rho_o + AT^m$, where $m = 3$ and $\rho_o$ is the residual impurity scattering part which is independent of temperature [22-24]. The fitted plots are depicted in Fig. 2 by red lines. The fitted power law plot is found to deviate from the experimental data at around 135 K and 49 K for $MgB_2$ and $AlB_2$ respectively. This yields $\Theta_D$ = 1350 K and 490 K for $MgB_2$ and $AlB_2$ respectively, within the assumption that the power law fitting deviates at $T = 0.1\Theta_D$. It is worth mentioning here that we followed the same procedure for fitting Bharathi et al [23] and Canfield et al [24]. Quantitatively, the $\rho(T)$ fitted $\Theta_D$ values *might not be correct*, but still give an idea that same is nearly double for $MgB_2$ than $AlB_2$. Fore more accurate determination experiments like heat capacity ($C_P$) and Raman spectroscopy need to be invoked. This will be discussed later, while reporting on $Cp(T)$ of $MgB_2$ in next sections.

Both *DC* and *AC* magnetic susceptibility measurements for $MgB_2$ are shown in Fig. 3. It is clear from both *DC* and *AC* susceptibility results that $MgB_2$ is a bulk superconductor with its critical transition temperature ($T_c^{onset}$) at 38.7 K. The $AlB_2$ sample did not exhibit any superconducting transition down to 2 K, and hence is concluded to be a non-superconductor. In *AC* susceptibility measurement, below the critical temperature, a sharp decrease in the real part of the AC susceptibility occurs, which reflects the diamagnetic shielding. In addition, below $T_c$ a peak appears in imaginary part of susceptibility (inset, upper panel Fig. 3) reflecting losses related to the flux penetrating inside the grains.

Fig. 4 shows the thermoelectric power variation $S(T)$ for $MgB_2$ and $AlB_2$ samples. Though thermoelectric power (*TEP*) for $MgB_2$ is positive, the same is negative for $AlB_2$. This indicates towards the hole type conductivity in $MgB_2$ system and electron type conductivity in $AlB_2$ system. Superconducting transition ($T_c$) is seen as $S=0$ at ~ 38 K, corroborating the $\rho(T)$ data The electronic structure of $MgB_2$ system involves two bands, i.e., $\sigma$ and the $\pi$-bands. It is believed that the $\sigma$-band is primarily responsible for the occurrence of the hole-type superconductivity in $MgB_2$. The electrons donated by Al tend to make the $\sigma$-band ineffective and thus only $\pi$-band contribution is there in $AlB_2$. It is clear that population of $\sigma$ -band is crucial in bringing about the superconductivity in various di-borides.

The temperature dependence of specific heat $C(T)$ is measured in zero field and for 14 Tesla applied on Quantum design PPMS. From the fitting of specific heat in normal state using Sommerfeld Debye expression:

$$C(T) = \gamma T + \beta T^3$$

We obtained the values of $\gamma$, Sommerfeld constant, and $\beta$ that gives the approximate value of Debye temperature. The values obtained are $\gamma=4.97325\pm0.1816$ mJ/mol-$K^2$, $\beta=0.01488\pm0.00009$ mJ/mol-$K^4$. These values are in good agreement with other reported values [18], though there are many contradictory reports also [14-17]. From the value of $\beta$ we calculated the value of Debye temperature using $\Theta_D = \left(\frac{234zR}{\beta}\right)^{\frac{1}{3}}$ (z being number of atoms per formula unit and R is gas constant) and found to be 731.9K. As mentioned in the resistivity section, the $\Theta_D$ value calculated from $Cp(T)$ is more authentic and the power law fitted resistivity is significantly overestimated as being nearly twice to the actual value. From the value of Sommerfeld constant we have calculated value of electronic Density of states at Fermi level $N(E_F)$ using formula $N(E_F) = \frac{3\gamma}{\pi^2 K_B^2}$ and is found to be 2.1086 states/eV per formula unit.

The jump in electronic specific heat that corresponds to superconducting transition is observed around 39K that agrees with our other measurements on thermo-electric power and resistivity. The specific heat data below $T_c$ is fitted using empirical $\alpha$-model [25], based on two discrete gaps $\Delta_1$ and $\Delta_2$ at $T=0$, both being closed at $T_c$.

According to α-model the ratio $\alpha=\Delta_0/k_BT_c$ is not fixed, but is assumed to be a fitting variable. The temperature dependence of the gap is however assumed according to *BCS* theory as $\Delta(t)=\Delta_0\delta(t)$, where $\delta(t)$ is normalised gap at reduced temperature $t=T/T_c$ and is used from the tables of Muhlschlegel [26].

We used two values of α (1.90 and 0.50) with weights of 0.35 and 0.65 to these respectively. This gave us reasonably good fitting to our experimental data and the values agree with earlier reports. The resulted fitted data is shown in Figure 1. The values of superconducting gaps obtained from the values of α are $\Delta_1=6.38$meV and $\Delta_2=1.59$meV. These values are in good agreement with other reports [27,28].

We also calculated electronic density of states (DOS) with both ELK and SIESTA code. Both the results match well with each other. The calculated and measured values of lattice parameters are given in Table 1 and are in good agreement with our experimental results. The numerical values of total and projected density of states (PDOS) are summarized in Table 2. From the value of *DOS* as obtained from specific heat measurements we calculated the electron-phonon coupling constant λ using relation

$$\frac{N_{expt}(E_F)}{N_{calc}(E_F)} = 1 + \lambda$$

Which gives λ=1.84, which corresponds to strong coupling *BCS* limits.

The electronic density of state for $MgB_2$ and $AlB_2$ are shown in Fig.6 and 7 respectively. The numerical values of total density of states and *l* and *m* resolved for various bands at Fermi level $N(E_F)$ are summarized in Table 2 which are in good agreement with recent calculations [11]. It can be noticed that for $MgB_2$ the majority of states come from σ and π 2p states of boron. The σ band consists of B-$2p_x$ and B-$2p_y$ orbitals that overlap quasi-two dimensionally in xy plane to form strong covalent bonds. π bands consists of weaker B-$2p_z$ orbital interactions. The $p_{olarized}$, which comes from *DZP*, is the polarisation of p orbitals in space. The σ bands (hole-type) are known to play an important role in electron-phonon coupling. In contrast for $AlB_2$ the boron π $2p_z$ and Al 3s and 3d states contribute significantly at the Fermi level. It is interesting to note the absence of σ band at Fermi level for $AlB_2$ leads to a weak electron phonon coupling and absence of superconductivity. Further in case of $AlB_2$ the boron π $2p_z$ and Al s and p domination makes the system electron type conductor instead of the boron σ band dominated hole type conductor $MgB_2$.

In summary, we inter compared physical and electronic properties of $MgB_2$ and $AlB_2$. An in depth analysis of heat capacity of $MgB_2$ is provided and it is concluded that the same is a strongly coupled two gap *BCS* superconductor.

Authors acknowledge the encouragement of their Director Prof. R.C. Budhani. JK acknowledges the financial support from Council of Scientific and Industrial Research (CSIR) India in terms of Senior Research Fellowship (SRF).

Table 1: Lattice parameters of $MgB_2$ and $AlB_2$ using GGA (PBEsol), the experimental values are given in the brackets.

| Sample | $a$ (Å) | $c/a$ | $T_c$ (K) |
|---|---|---|---|
| $MgB_2$ | 3.10 (3.08) | 1.13 (1.14) | ~38 |
| $AlB_2$ | 3.02 (3.01) | 1.09 1.06 | NSC |

Table 2: Percentage contribution to electronic density of states at Fermi level $N(E_F)$ (states/eV/cell/spin) from different orbitals

| Sample | Total DOS | $B-2p_x$ | $B-2p_y$ | $B-2p_z$ | Mg/Al-3s | Mg/Al-3p | Mg/Al-3d | $B-2p$ polarized |
|---|---|---|---|---|---|---|---|---|
| $MgB_2$ | 0.3696 | 0.0689 | 0.0689 | 0.1606 | 0.0155 | 0.0132 | 0.0000 | 0.0403 |
| $AlB_2$ | 0.1792 | 0.0035 | 0.0035 | 0.0464 | 0.0343 | 0.0142 | 0.0337 | 0.0305 |

**Figure Captions:**

Figure.1 Rietveld refined plots for (a) $MgB_2$ and (b) $AlB_2$ samples. X-ray experimental diagram (red dots), calculated pattern (black continuous line), difference (lower blue continuous line) and calculated Bragg position (vertical green lines in middle).

Figure.2 Resistivity variation with temperature $\rho(T)$ of $MgB_2$ and $AlB_2$, red lines are the fitted plot according to power law.

Figure.3 *DC* and *AC* magnetization of $MgB_2$ system

Figure.4 Thermoelectric power vs. temperature $S(T)$ plots for $MgB_2$ and $AlB_2$ samples

Figure.5 Two gap fitting of electronic specific heat below $T_c$ for $MgB_2$ using $\alpha$-model, the inset shows the individual contributions from the two superconducting gaps, as mentioned in the text.

Figure.6 Calculated total and partial electronic density of states for $MgB_2$

Figure.7 Calculated total and partial electronic density of states for $AlB_2$

Fig. 1

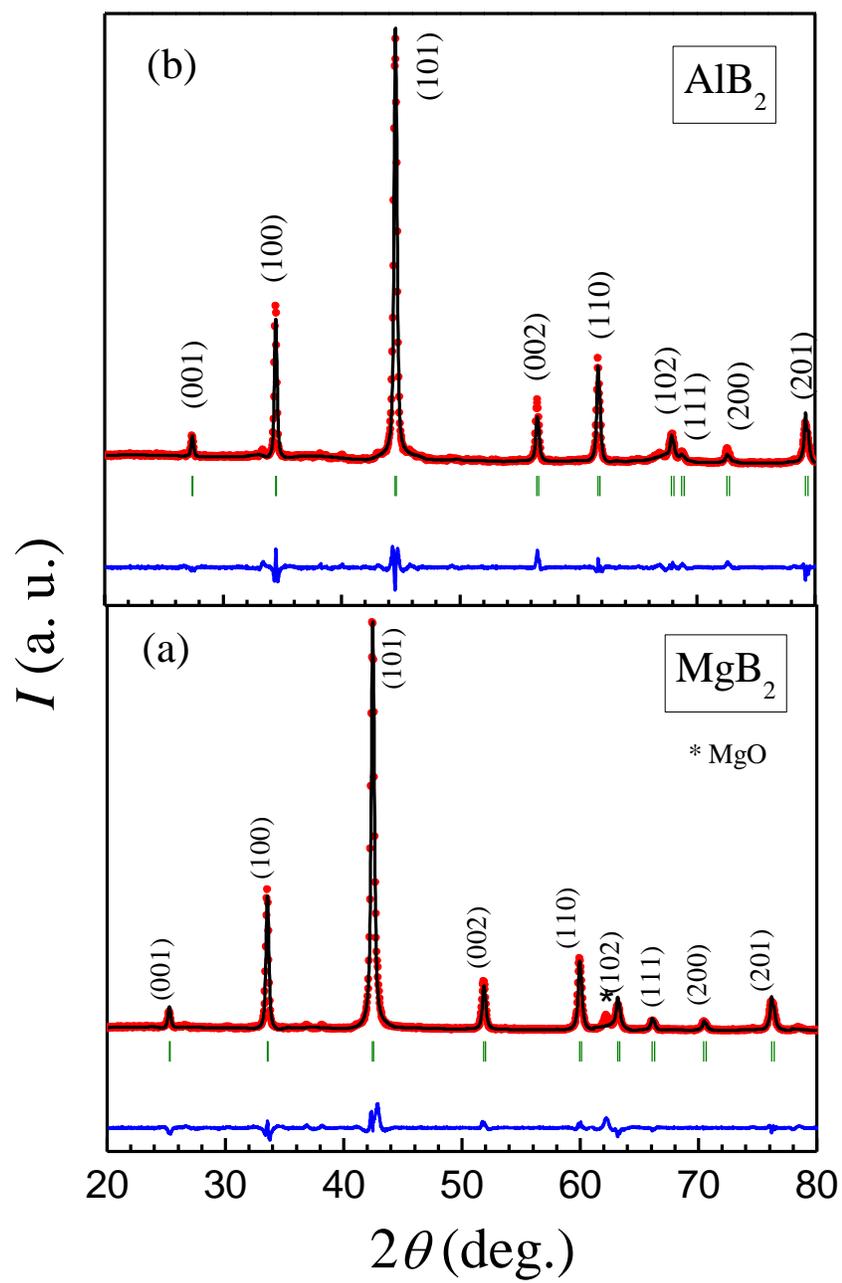

Fig.2

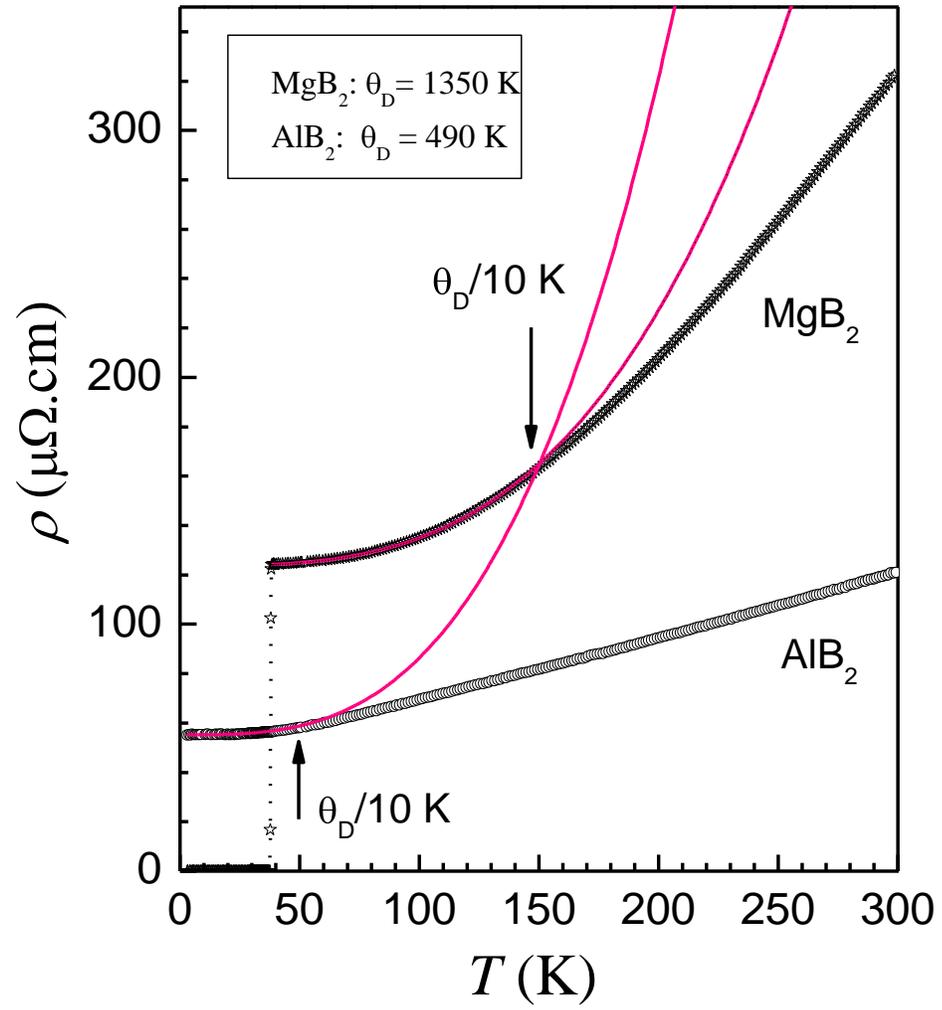

Fig. 3

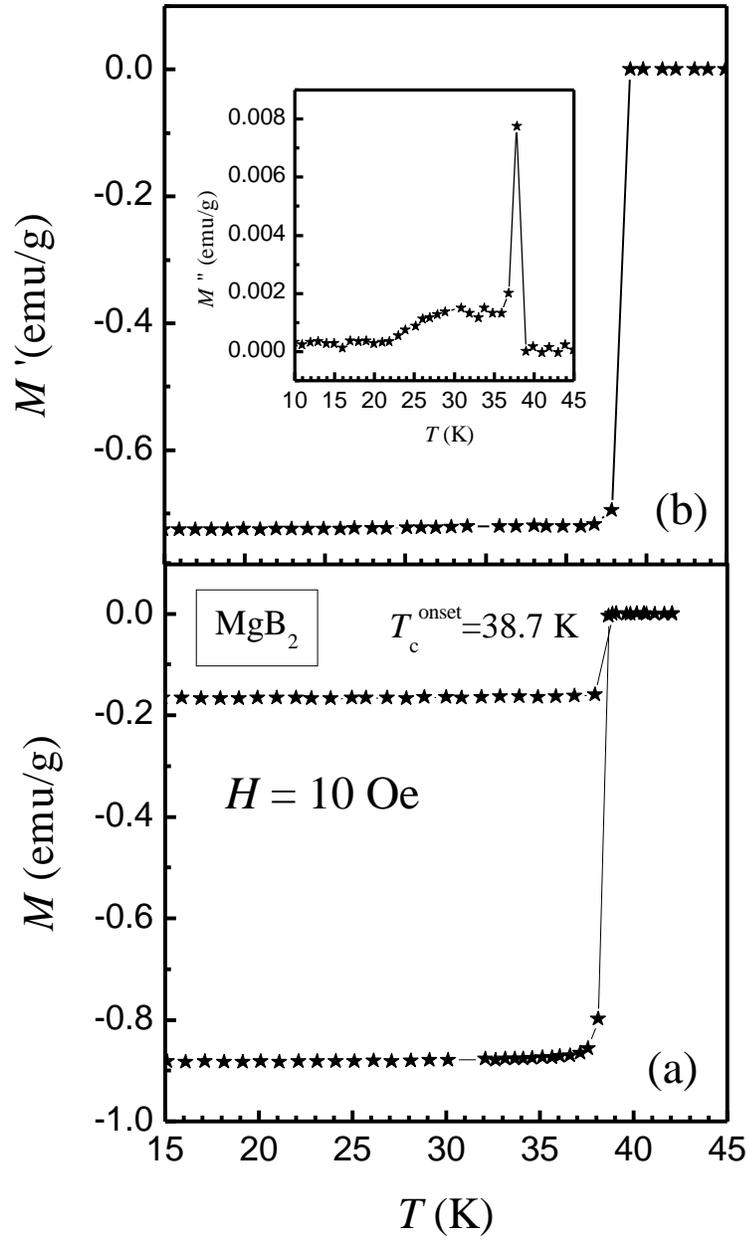

Fig. 4

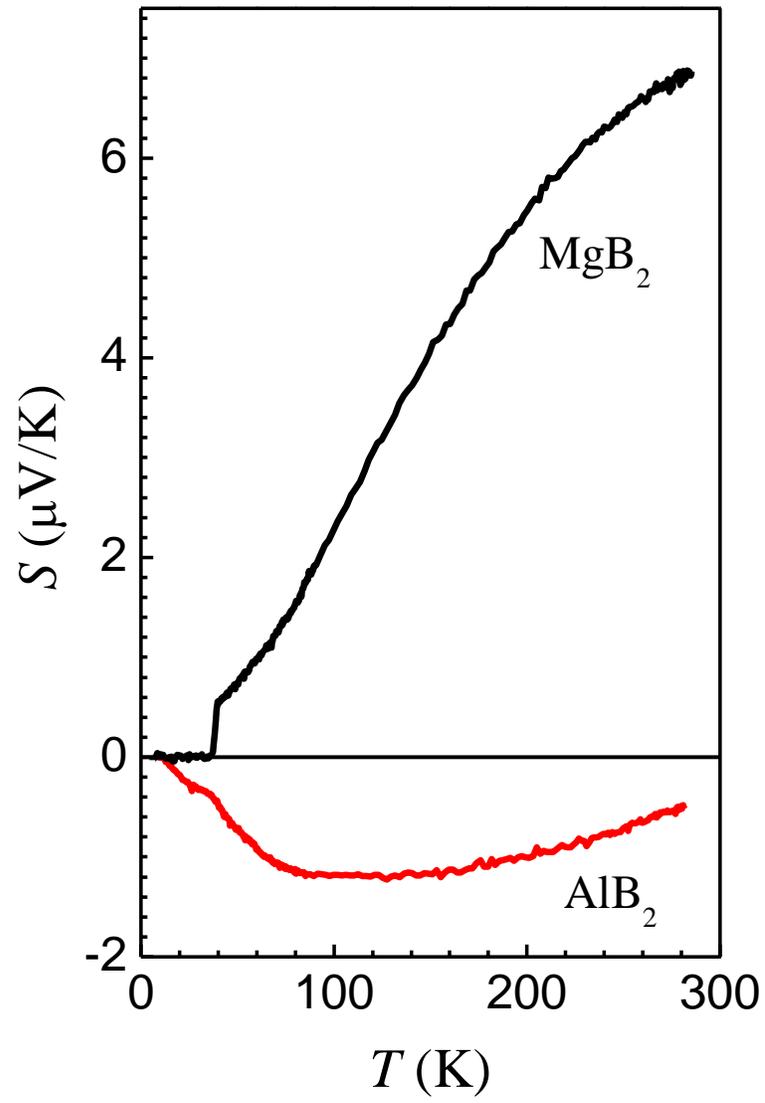

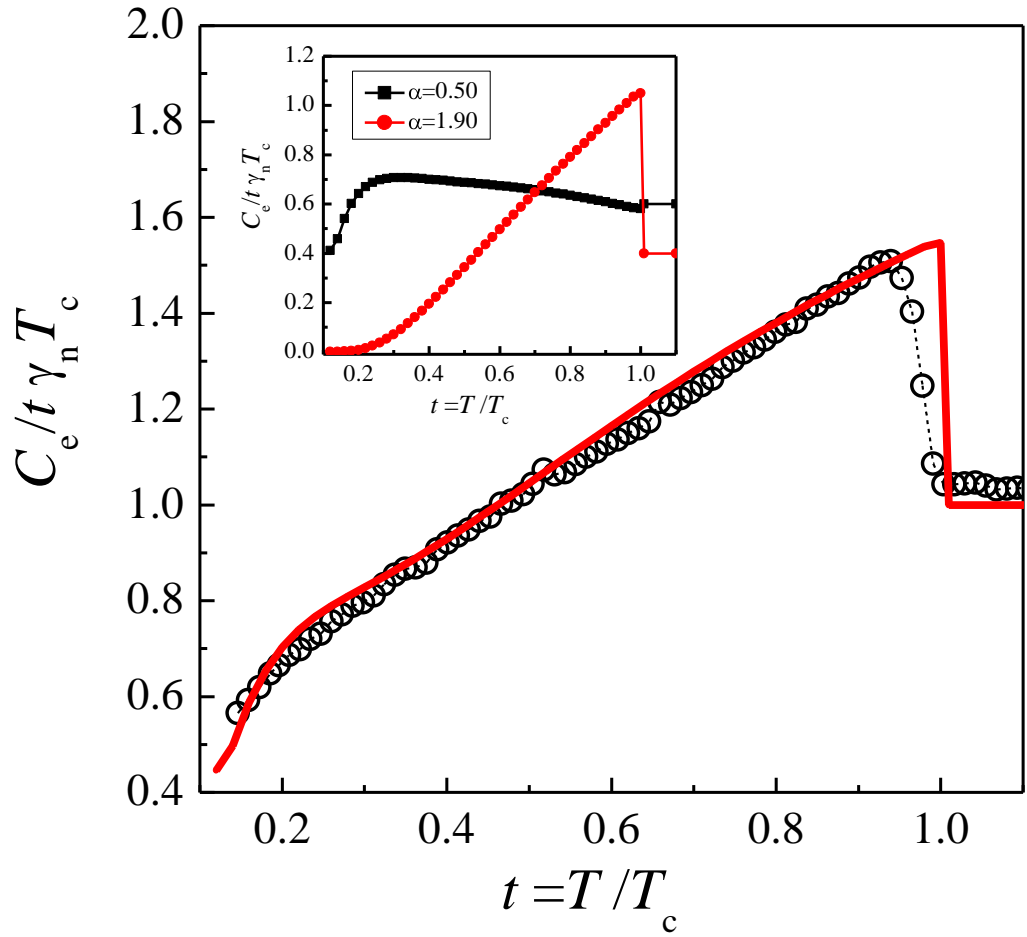

Fig. 6

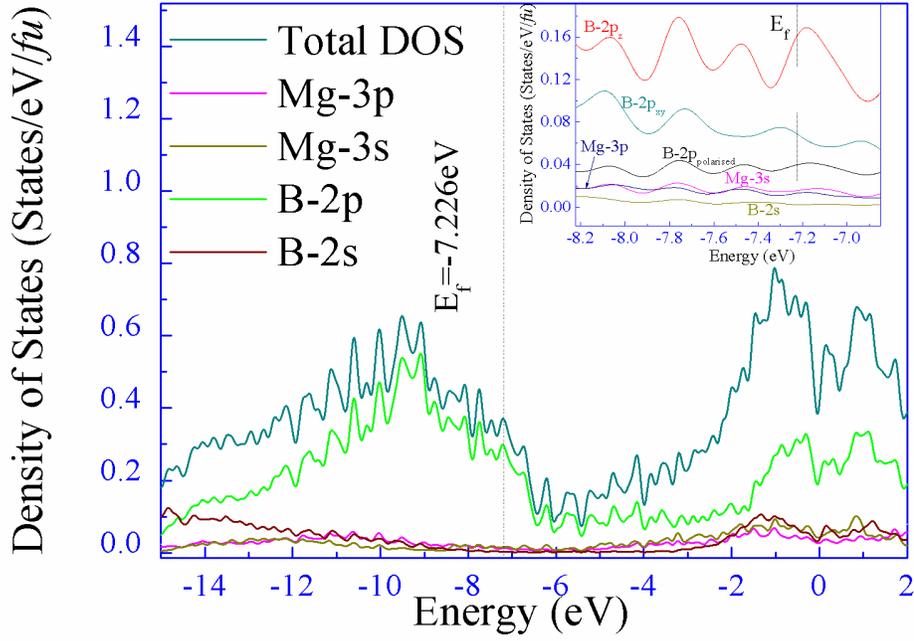

Fig. 7

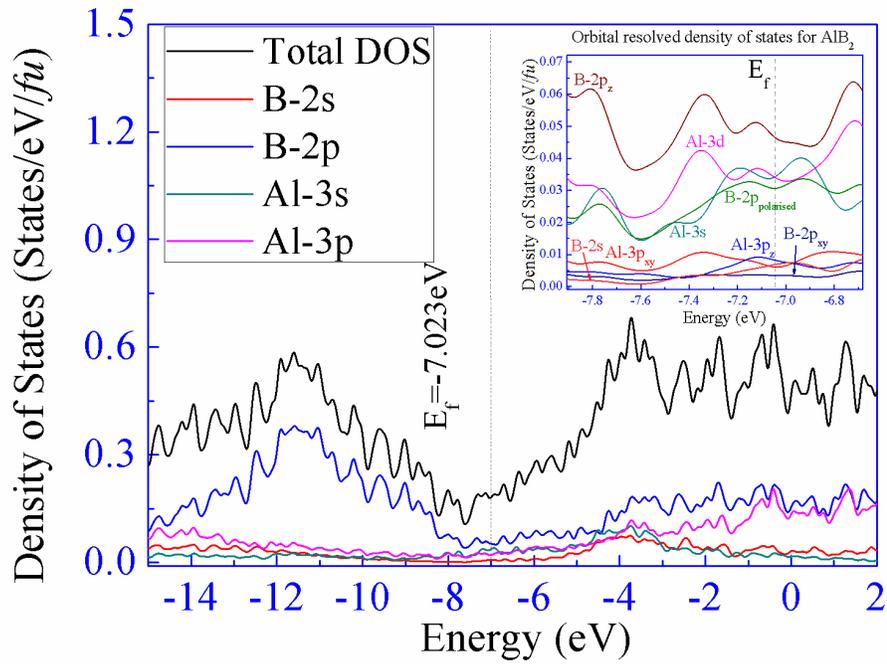

16